# Thermal conductivity reduction by 60° shuffle-set dislocation arrays embedded in silicon nano-films

Zhimin Jiang,[†] Haisheng Fang,[*,†] Zhenya Lv,[†] Xiangang Luo,[†] Lili Zheng[*,‡]

[†]State Key Laboratory of Coal Combustion, School of Energy and Power Engineering, Huazhong University of Science and Technology, Wuhan 430074, Hubei, P. R. China

[‡]School of Aerospace Engineering, Tsinghua University, Beijing, P. R. China

**Abstract**

Based on the Debye-Callaway and the Klemens model, as well as molecular dynamics, the paper proposes mechanism of thermal conductivity reduction by embedding dense 60° shuffle-set dislocation arrays into silicon nano-films. Thermal conductivity reduction to 2% of that of bulk silicon has been obtained. The reduction is found mainly due to longitudinal phonon scattering at the dislocation cores, where the scattering rate is stronger than that presented by Klemens. Within an effective diameter of about 9 nm around their cores, the dislocations locally scatter phonons, resulting in a dramatical density-dependent reduction of thermal conductivity for a dislocation density larger than $10^{14}$ m$^{-2}$.

Thermal conductivity is a key thermophysical property affecting applications of materials. For example, thermoelectric (TE) performance of a material is evaluated by the dimensionless figure of merit [$ZT=S^2\sigma T/(\kappa_{electron}+\kappa_{lattice})$], depending on electrical conductivity ($\sigma$), Seebeck coefficient ($S$) and thermal conductivity ($\kappa$).[1] Low thermal conductivity and large power factor ($S^2\sigma$) help to improve efficiency of converting thermal energy into electricity. In jet engine industry, nanoscale low-$\kappa$ and high-strength heat insulators are required to reduce temperature rapidly as a protection of core parts.[2, 3] As an illustrative example, thermal-regulated nano-silicon shows the way for designing low-$\kappa$ materials but maintaining good electrical and mechanical properties.[4, 5] Deterioration of thermal transport in nanoscale is attributed to strong phonon scattering on boundaries.[6, 7] Further investigations point out that appropriate surface coatings or increasing surface roughness can enhance the boundary scattering and reduce thermal conductivity effectively.[4, 8, 9] Thus, the $\kappa$ reduction due to the conventional scattering mechanisms, such as boundary, isotope, doping and dislocation, has not reached its limit yet.

Dislocation scattering belongs to conventional mechanisms of thermal conductivity reduction.[10, 11] Considering its speciality in regulating thermal and electrical properties as discussed in the following, deep excavations of the phonon scattering by dislocations is extremely necessary. In TE material design, the power factor could be maintained or even enhanced as dislocations appear.[12, 13] For example, dense dislocation arrays embedded in bismuth-antimony-telluride grain boundaries make $\kappa_{lattice}$ reduce to 0.33 W/(mK),[12] and enhance *ZT* to 1.86 at 300 K as $\kappa_{electron}$ is negligible in semiconductors ($\kappa=\kappa_{lattice}$).[12] Experimental research further indicates that confining dislocations in a small region would enhance electrical properties.[13] However, possibility of dislocation movement is proportional to *exp(-Q/k$_B$T)*, leading to annihilation of dislocations with the reverse Burgers' vectors at high temperatures.[14, 15] Although a high activation energy, *Q* (about 2 eV), in covalent materials would decrease the possibility,[14] the degeneration of performance is still unavoidable due to dislocation motion. Fortunately, recent investigations of silicon dislocation disclose that 60° shuffle-set dislocations with reconstructed cores are sessile, which do not move even at high temperatures.[16, 17] The type of dislocations could be fabricated by applying a high confining pressure,[18, 19] making it possible to embed dense dislocations in nanostructures without annihilations at high working temperatures. Therefore, more attentions should be put on the impacts of dense 60° silicon shuffle-set dislocations on the thermal transport in nanostructures.

---
[*]Corresponding author. Email: hafang@hust.edu.cn (Haisheng Fang). Phone: +86 027-87542618. E-mail: zhenglili@tsinghua.edu.cn (Lili Zheng). Phone: +86 010-62797961.

Thermal transport along and across dislocations in bulk materials is first investigated by Klemens using the perturbation theory and the Debye-Callaway model.[11] Nevertheless, the dislocation core scattering term in Klemens' formulas is treated simply as a line of vacancy defects,[11] which have not been approved because of lacking data. In nanostructures, boundary scattering term of the Callaway model needs to be treated carefully due to the strong structure dependence of the phonon mean free path.[7, 20] Recently, investigations about dislocation effects on the nanostructures mainly focus on nanowires with screw dislocations,[21, 22] leading to a lack of research regarding nano-films coupled with the specific 60° shuffle-set dislocation. In this paper, we firstly propose a scheme of embedding the 60° shuffle-set dislocations with sessile cores as an array in silicon nano-films. The Callaway's and the Klemens' models, as well as molecular dynamics (MD) simulations, are then adopted to understand how a dislocation array makes thermal transport deteriorated. Averaged phonon scattering rates of dislocations, local phonon density of states (LPDOS) and local temperature distributions are computed to uncover anharmonic phonon-phonon scattering around the dislocations.

The MD simulation scheme performed by LAMMPS package[23] is shown in Fig. 1 with 48 [1 1 1] atom layers in the z direction (about 14.8 nm). Periodic boundaries are adopted in the x and y directions, making the system act approximately as an infinite nano-film. Simulation size in y direction is set to 3.08 nm to achieve a balance between simulation accuracy and computational load (see Fig. S4 in Supporting Information). The size along the x axis (width) changes from 3.84 nm to 18.48 nm, and the dislocation density in the array accordingly varies from $2.68\times10^{16}$ to $5.58\times10^{15}$ m$^{-2}$. Thermal conductivity across the film is computed by Non-equilibrium MD method with the Stillinger-Weber (SW) potential (see Supporting Information).[24] The potential has been successfully applied to investigate thermal properties of both nanostructure and bulk silicons.[6, 25] After the simulations, the parameters of an improved Debye-Callaway model for bulk silicon are proposed to fit the published data (see Fig. S5), and then the boundary scattering term is modified to fit the simulation results of perfect nano-films (Fig. S4). The Klemens expressions of dislocations are adopted to analyze the possible mechanisms for reduction of thermal conductivity in dislocation-embedded films.[11, 26] The adopted expressions for Callaway's and Klemens' models are listed in Table I with the corresponding parameters given in Table II and Table III (see Supporting Information for a detailed description).

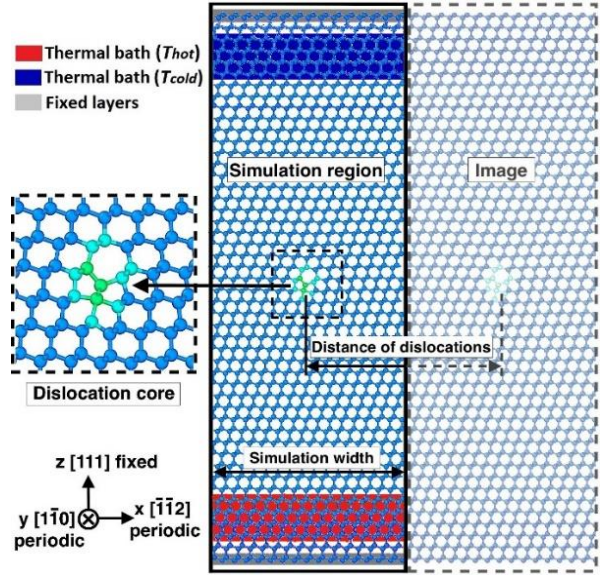

Fig. 1. MD scheme for calculation of thermal conductivity. The 60° shuffle-set dislocation with a sessile core is shown on the left, which is reconstructed from an unstable 60° shuffle-set pristine dislocation.[16, 27] Due to periodic boundaries, dislocations lie in a line to become an array. In order to eliminate the effect of localized edge mode of phonons, one free layer is set between the thermal bath and the fixed layer.[28, 29]

Table I. The adopted expressions for Callaway's and Klemens' models.

| | |
|---|---|
| Thermal conductivity[26, 30, 31] | $\kappa = \kappa_1 + \kappa_2$ $\kappa_2 = \frac{1}{3} \sum_n \frac{k_B}{2\pi^2 \upsilon_n} \left(\frac{k_B T}{\hbar}\right)^3 \frac{\left[\int_0^{\Theta_n^i/T} \tau_{c,n}(x) \tau_{N,n}(x)^{-1} x^4 e^x (e^x-1)^{-2} dx\right]^2}{\int_0^{\Theta_n^i/T} \tau_{c,n}(x) \left[\tau_{N,n}(x) \tau_{u,n}(x)\right]^{-1} x^4 e^x (e^x-1)^{-2} dx}$ $\kappa_1 = \frac{1}{3} \sum_n \frac{k_B}{2\pi^2 \upsilon_n} \left(\frac{k_B T}{\hbar}\right)^3 \int_0^{\Theta_n/T} \tau_{c,n}(x) \frac{x^4 e^x}{(e^x-1)^2} dx$ |
| Combined scattering rate (SR) | $\tau_c^{-1} = \tau_N^{-1} + \tau_u^{-1}$ |
| Unconserved-momentum SR[11, 26] | $\tau_{u,n}^{-1} = \tau_{U,n}^{-1} + \tau_{SD,n}^{-1} + \tau_{ED,n}^{-1} + \tau_{DC,n}^{-1} + \tau_{B,n}^{-1}$ |
| Umklapp SR[32-34] | $\tau_{U,n}^{-1} \approx \frac{\hbar \gamma_n^2}{M \upsilon_n^2 \Theta_n} \omega^2 T \exp\left(\frac{-\Theta_n}{3T}\right)$ |
| Normal process SR[12] | $\tau_{N,n}^{-1} \approx \beta \times \tau_{U,n}^{i\ -1}$ |
| Boundary SR[12, 26] | $\tau_{B,n}^{-1} = \frac{\upsilon_n}{L}$ |
| Screw dislocation SR[11] | $\tau_{SD,n}^{-1} = A \times N_D B_{SD}^2 \gamma_n^2 \omega$ |
| Edge dislocation SR[11] | $\tau_{ED,n}^{-1} = A \times N_D B_{ED}^2 \gamma_n^2 \omega \left\{\frac{1}{2} + \frac{1}{24}\left(\frac{1-2v}{1-v}\right)^2 \left[1 + \sqrt{2}\left(\frac{\upsilon_L}{\upsilon_T}\right)^2\right]^2\right\}$ |

| Dislocation core SR[11, 12] | $\tau_{DC,n}^{-1} = N_D \times \frac{3V_0^{4/3}\omega^3}{4\upsilon_n^2} \times \left(\frac{\Delta M}{M}\right)^2$ |
|---|---|

*The subscript n represents three acoustical phonon branches, two transverse (T) branches and one longitudinal (L) branch. $\tau_{SD}$ and $\tau_{ED}$ are the scattering rates outside the dislocation cores due to the strain.

Table II. Parameters of the bulk isotopic silicon.[35, 36]

| Atomic mass | $M$=4.66×10$^{-26}$ kg |
|---|---|
| Gruneisen anharmonicity parameter | $\gamma_L$=1.09, $\gamma_T$=0.45 |
| Group velocity | $\upsilon_L$=8430 m/s, $\upsilon_T$=5840 m/s |
| Constant for Normal process | $\beta$=1.5 |
| Characteristic length | $L$=14 mm (bulk) |
| Debye temperature | $\Theta_L$=566 K, $\Theta_T$=216 K |

Table III. Parameters for predicting $\kappa$ of the dislocation-embedded nano-films.

| Characteristic length | $L$=43 nm (silicon film) |
|---|---|
| Constant for dislocation | $A$=1.1 |
| Dislocation Density | $N_D$=1/($L_{21}$×$D$) |
| Film thickness | $L_{21}$=9.7 nm |
| Distance between dislocations | $D$>3.8 nm |
| Atomic volume | $V_0$=2.0×10$^{-29}$ m$^3$ |
| Poison's ratio | $v$=0.27 |
| Mass deviation in the core | $\Delta M$=1.35$M$ |
| Burgers' vector | $B$=1/2[110]×$L_{lattice}$ |
| Screw component of $B$ | $B_{SD}$=(|$B$|$^2$cos(60°))$^{1/2}$ |
| Edge component of $B$ | $B_{ED}$=(|$B$|$^2$sin(60°))$^{1/2}$ |
| Lattice constant | $L_{lattice}$=0.543 nm |

*Since 60° shuffle-set dislocations are mixed dislocations, the Burgers' vector is decoupled into the screw component and the edge component to calculate the scattering rate. $L$ is determined to be about 43 nm by fitting the simulation results of perfect films.

The thermal conductivities of 60° shuffle-set dislocation-embedded nano-films are presented in Fig.2. The MD simulation results show a remarkable reduction of $\kappa$ with a decreasing distance between two adjacent dislocations in the array. As the distance changes from 18.48 nm to 3.83 nm, $\kappa$ drops from 12.39 W/(mK) to 6.46 W/(mK), less than that of the perfect nano-film (18.0 W/(mK), see Fig. S4). When the distance decreases to 2 nm, as predicted by the curve with $\Delta M/M$=1.35, $\kappa$ goes down to about 4.5 W/(mK), about 25% of the original $\kappa$ of the thin film and about 2% of the bulk.[37] It is reasonable to foretell that further decrease of the film thickness or embedding multiple arrays in the film, $\kappa$ of the film would further drop. Besides, in Klemens' model, a dislocation core is treated as a line of vacancy defects, leading to mass deviation $\Delta M$=$M$. But for 60° shuffle-set dislocations in silicon, such an unreliable treatment obtains much larger values than MD does as shown in Fig. 2. Using $\Delta M/M$=1.35, reasonable predictions can be achieved, which means that the 60° shuffle-set dislocation cores play as much stronger phonon scattering sources than that proposed by Klemens. The calculated $\kappa$ of films with the pristine 60° shuffle-set dislocation in Fig.2 further illustrates that the increased $\Delta M/M$ is not because of the core reconstruction. We ascribe it to the diamond lattice structure of silicon, which generates larger Burgers' vectors of dislocations than other lattice structures. With a large vector, the dislocation core in silicon behaves as more than one line of vacancy, which has been ignored by Klemens.

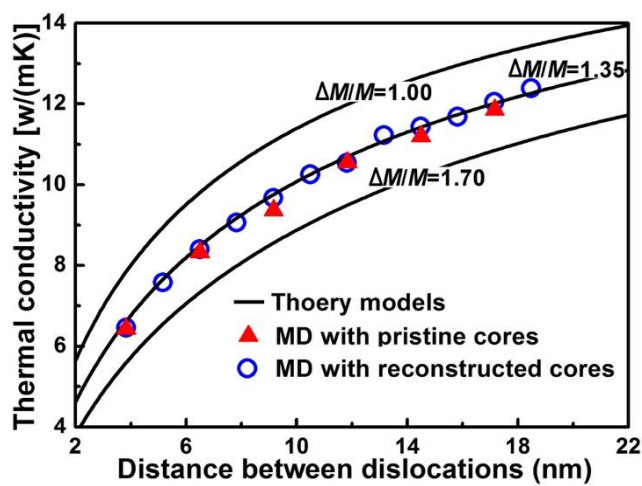

Fig. 2 $\kappa$ of nano-films with 60° shuffle-set dislocations embedded. The red triangles and the blue circles stand for $\kappa$ calculated by MD simulations with pristine (unreconstructed) cores and with reconstructed cores, respectively, and the black solid lines represent $\kappa$ predicted by Callaway's and Klemens' models with different $\Delta M/M$.[11, 12]

With expressions and parameters in Tables I, II and III, thermal conductivities of the nano-film and the bulk silicon depending on dislocation densities are calculated (See Fig. 3). Only when the dislocation density reaches $10^{14}$ m$^{-2}$, $\kappa$ reduction becomes distinct. A large reduction of $\kappa$ is observed in both bulk and nano-film silicon when the density become larger. With the density of $10^{17}$ m$^{-2}$, the $\kappa$ of bulk silicon decreases to 5.3 W/(mK), comparable to the $\kappa$ of short silicon nanowires.[6] If embedding dislocations as an array in the nano-film, it would further decrease to 3 W/(mK) with the same density, nearly 43% smaller than that of dislocation-embedded bulk silicon. The significant reduction of $\kappa$ in dislocation-embedded nano-films due to coupling of dislocation and boundary scattering illustrates a potential direction for the future design of thermoelectric materials and heat barriers.

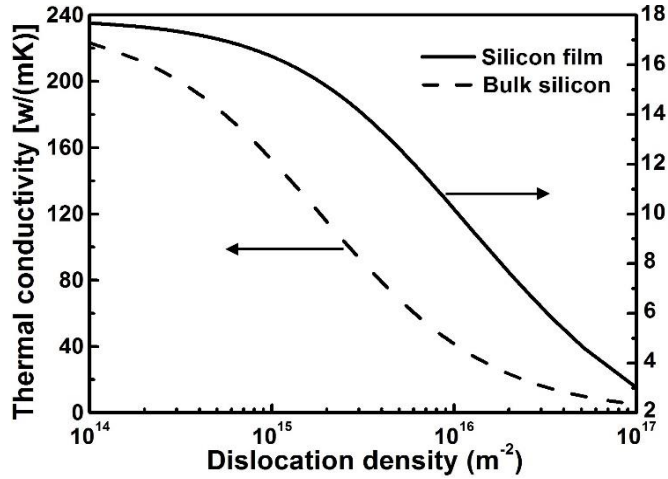

Fig. 3. Thermal conductivities of the bulk and the nano-film silicons at 300 K with different dislocation densities. The κ of the bulk and the nano-film silicons without dislocations at 300 K are 237 W/(mK) and 18.0 W/(mK) (see Fig. S4 and Fig. S5).

Although, as described by Klemens, small distance between two dislocations can reinforce dislocation strain field, and results in an enhancement of phonon scattering.[12, 38] The high consistency in Fig. 2 indicates the reinforcement in the currently investigated system is not strong enough to affect κ. Figure 4 presents averaged phonon scattering rates computed from the formulas in Table I, accounting for the unobservable reinforcement. Obviously, the longitudinal and the transverse phonon scattering rates at the dislocation cores have nearly three times larger than those caused by stress fields. As a result, κ reduction is dominated by dislocation-core phonon scattering, especially by the longitudinal phonon scattering at the core, while scatterings caused by the strain field of dislocations play a minor role in κ reduction. It is coincided with the calculated local thermal resistances, which almost keeps unchanged outside a screw dislocation core.[22] Additionally, from Fig. 2, the smaller the dislocation distance is, the faster the κ decreases, due to more phonons scattered by dislocation cores.

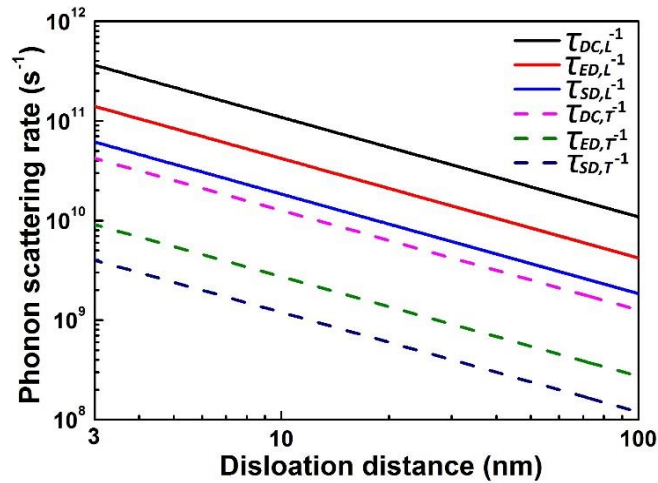

Fig. 4. The phonon scattering rates calculated from the formulas in Table I. The dashed and the solid lines represent the transverse and longitudinal phonon scattering rates, respectively.

For further investigation of phonon scattering, local phonon density of states (LPDOS), as a density distribution function of phonon eigenstates of each atom, is calculated by MD and Welch method (see Supporting information).[39] Its consistence degree between the atoms represents the smooth degree of phonon transport. Figure 5a shows the LPDOS of one atom line along the y axis with the same (x, z) = (-0.79, -0.20). The high consistency of the LPDOS implies that phonons with the wave vectors along the dislocation line (the y direction) are seldom affected by the dislocation strain field. In contrast, as shown in Fig. 5b, the large inconsistency of the LPDOS at the core region illustrates that transport of phonons with the wave vectors along x or z directions would be dramatically affected by the dislocation core. As a result, phonons oblique to the dislocation line will be scattered, which means thermal transport across the dislocation array would deteriorate more than that along the array. The result coincides with theoretical analysis by Klemens.[11] Besides, the inconsistency is larger at the dislocation core than that at the peripheral region as given in Fig. 5c, demonstrating that the core is a stronger scattering source than the peripheral region. It accounts for the fast reduction of $\kappa$ by a dense dislocation array as mentioned above. Atoms at the core with different neighboring structures have special local vibrational modes, thus making thermal transport across the dislocation core extremely difficult.

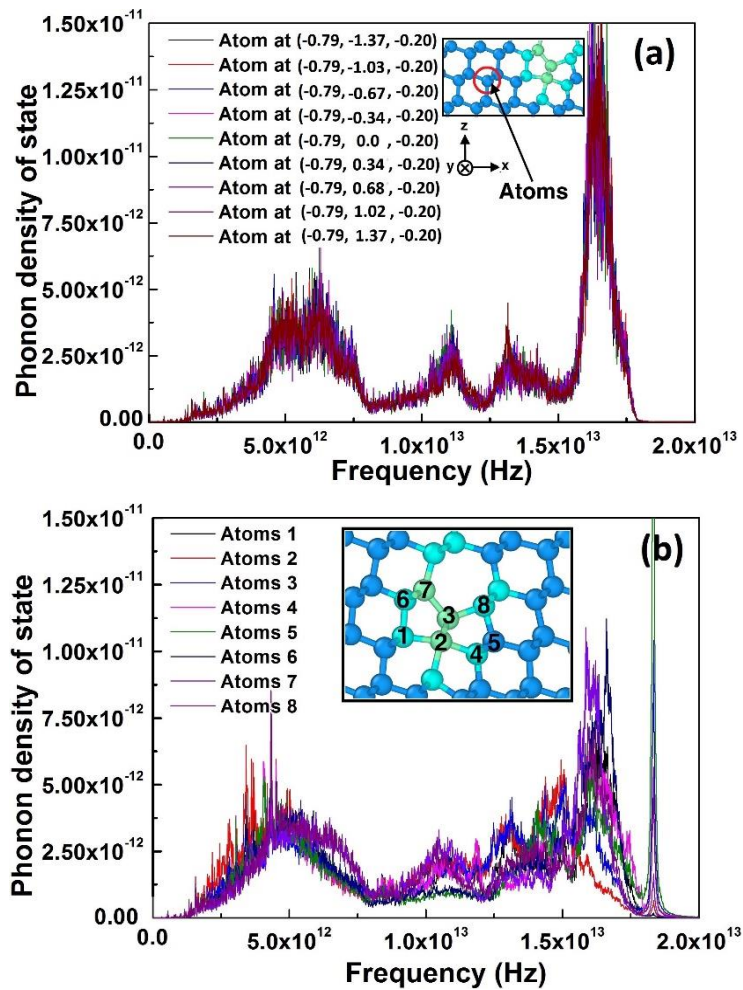

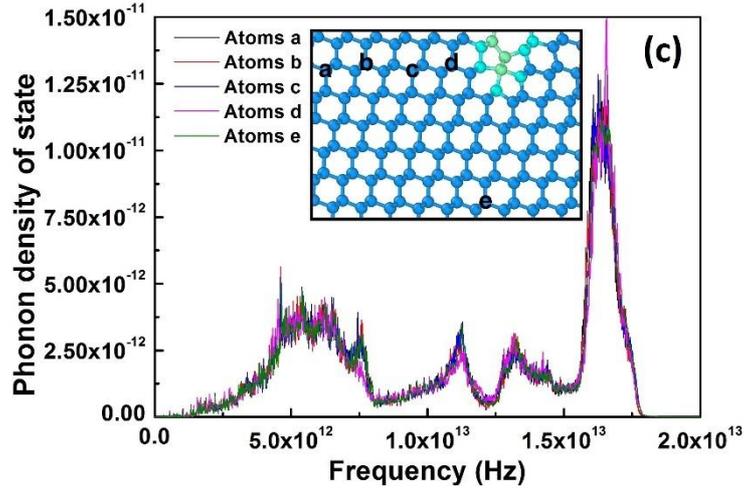

Fig. 5 (a) The LPDOS of a line of atoms beside the core, (b) the high inconsistency of LPDOS of atoms at the core, and (c) the calculated LPDOS of the peripheral region around the core.

To reveal the influence degree at the regions affected by the dislocations, the atomic temperature distribution, as shown in Fig. 6a and 6b, is calculated when a temperature gradient is applied. In lattice, heat flux $J$ can be approximately written as

$$J = -\frac{1}{3}\upsilon^2 \tau C_V \nabla T, \qquad (1)$$

where $\upsilon$, $\tau$ and $C_V$ are the averaged phonon-group velocity, the relaxation time and the specific heat capacity, respectively.[40] Assuming that $\upsilon$ and $C_V$ are constants, it is obvious that $\tau^{-1}$ (phonon scattering rate) is proportional to $\nabla T/J$. Further assuming that heat flux is constant across the film in Fig. 6a (such assumption is not rigid close to the core region), the uniform temperature gradient obtained over the region A and C (see Fig. 4a) tells that phonon scattering rates of these regions are not affected by the dislocation, where $\tau^{-1}$ is only determined by the Umklapp and the Normal process. Thus, the temperature gradient in Fig. 6a reveals that the effective diameter of the region, where dislocation contributes to phonon scattering, is only about 9 nm, revealing that a large reduction in $\kappa$ could only be obtained by dense dislocations as predicted in Fig. 3. Figure 6b clearly shows that the dislocation blocking the phonon transport results in a large temperature gradient at the core due to extreme inconsistence of LPDOS as revealed in Fig. 5b.

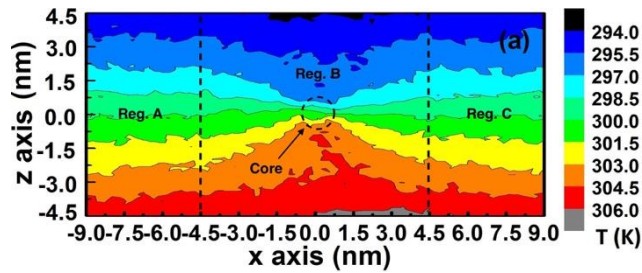

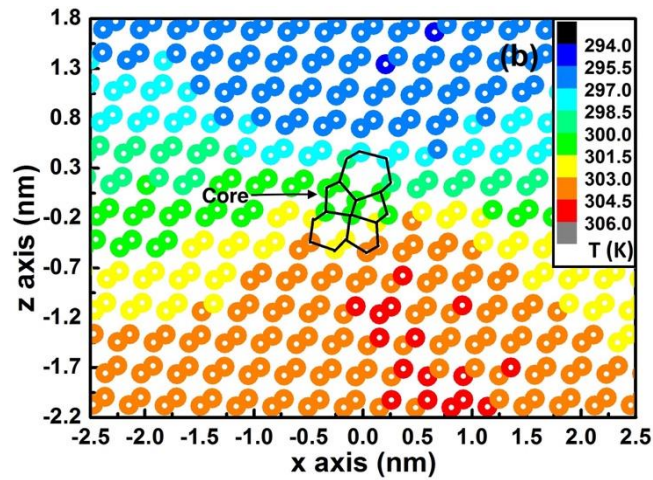

Fig. 6 (a) The temperature distribution when applying a temperature gradient, and (b) the local temperature distribution of atoms around the dislocation core.

**Conclusion**

Using the Debye-Callaway and the Klemens models, as well as MD simulations, the paper illustrates thermal transport through 60° shuffle-set dislocation arrays embedded in silicon nano-film. A large $\kappa$ reduction has been observed due to coupling phonon scatterings by boundaries and dislocations. It is further found that dislocations only locally scatter phonons within an effective diameter of about 9 nm, and that the dislocation cores in silicon plays as scattering sources much stronger than that predicted by Klemens. The theory and the LPDOS analysis disclose the reduction is mainly controlled by longitudinal phonon scattering at the dislocation cores, where inconsistency of LPDOS due to complex core structure is responsible for the scattering. The small effective diameter leads to a strong thermal conductivity reduction depending on dislocation density, especially when it is larger than $10^{14}$ m$^{-2}$. Moreover, theoretical analysis discloses that by appropriately manipulating dense dislocations to be an array embedded in nano-films, thermal transport could get 98% smaller than that of bulk silicon. It is an important implication for designing high-performance thermoelectric materials or heat barriers.

**Acknowledgements**
The work is supported by National Natural Science Foundation of China (No.51476068). Numerical calculation presented is partially carried out at the High Performance Computing Center (http://grid.hust.edu.cn/hpcc).

**References**
1. Venkatasubramanian, R.; Siivola, E.; Colpitts, T.; O'quinn, B. *Nature* **2001,** 413, (6856), 597-602.
2. Clarke, D. R.; Phillpot, S. R. *Materials Today* **2005,** 8, (6), 22-29.
3. Miller, R. A. *Journal of thermal spray technology* **1997,** 6, (1), 35-42.
4. Hochbaum, A. I.; Chen, R.; Delgado, R. D.; Liang, W.; Garnett, E. C.; Najarian, M.;

Majumdar, A.; Yang, P. *Nature* **2008,** 451, (7175), 163-167.
5. Boukai, A. I.; Bunimovich, Y.; Tahir-Kheli, J.; Yu, J.-K.; Goddard Iii, W. A.; Heath, J. R. *Nature* **2008,** 451, (7175), 168-171.
6. Yang, N.; Zhang, G.; Li, B. *Nano Today* **2010,** 5, (2), 85-90.
7. Mingo, N.; Yang, L.; Li, D.; Majumdar, A. *Nano Letters* **2003,** 3, (12), 1713-1716.
8. Martin, P.; Aksamija, Z.; Pop, E.; Ravaioli, U. *Physical review letters* **2009,** 102, (12), 125503.
9. Sansoz, F. *Nano letters* **2011,** 11, (12), 5378-5382.
10. Mott, N. *The London, Edinburgh, and Dublin Philosophical Magazine and Journal of Science* **1952,** 43, (346), 1151-1178.
11. Klemens, P. *Proceedings of the Physical Society. Section A* **1955,** 68, (12), 1113.
12. Kim, S. I.; Lee, K. H.; Mun, H. A.; Kim, H. S.; Hwang, S. W.; Roh, J. W.; Yang, D. J.; Shin, W. H.; Li, X. S.; Lee, Y. H. *Science* **2015,** 348, (6230), 109-114.
13. Bennett, N. S.; Byrne, D.; Cowley, A. *Applied Physics Letters* **2015,** 107, (1), 013903.
14. Rabier, J.; Pizzagalli, L.; Demenet, J. *Dislocations in solids* **2010,** 16, 47-108.
15. Chaudhuri, A.; Patel, J.; Rubin, L. *Journal of Applied Physics* **1962,** 33, (9), 2736-2746.
16. Pizzagalli, L.; Godet, J.; Brochard, S. *Physical review letters* **2009,** 103, (6), 065505.
17. Asaoka, K.; Umeda, T.; Arai, S.; Saka, H. *Materials Science and Engineering: A* **2005,** 400, 93-96.
18. Yonenaga, I. *Engineering Fracture Mechanics* **2015,** 147, 468-479.
19. Rabier, J.; Cordier, P.; Demenet, J.; Garem, H. *Materials Science and Engineering: A* **2001,** 309, 74-77.
20. Xie, G.; Guo, Y.; Li, B.; Yang, L.; Zhang, K.; Tang, M.; Zhang, G. *Phys. Chem. Chem. Phys.* **2013**.
21. Xiong, S.; Ma, J.; Volz, S.; Dumitrică, T. *Small* **2014,** 10, (9), 1756-1760.
22. Ni, Y.; Xiong, S.; Volz, S.; Dumitrică, T. *Physical review letters* **2014,** 113, (12), 124301.
23. Plimpton, S. *Journal of computational physics* **1995,** 117, (1), 1-19.
24. Stillinger, F. H.; Weber, T. A. *Physical review B* **1985,** 31, (8), 5262.
25. Volz, S. G.; Chen, G. *Applied Physics Letters* **1999,** 75, (14), 2056-2058.
26. Callaway, J. *Physical Review* **1959,** 113, (4), 1046.
27. Jiang, Z.; Fang, H.; Luo, X.; Xu, J. *Engineering Fracture Mechanics* **2016,** 157, 11-25.
28. Jiang, J.-W.; Chen, J.; Wang, J.-S.; Li, B. *Physical Review B* **2009,** 80, (5), 052301.
29. Chen, J.; Zhang, G.; Li, B. *Journal of the Physical Society of Japan* **2010,** 79, (7), 074604.
30. Zou, J.; Kotchetkov, D.; Balandin, A.; Florescu, D.; Pollak, F. H. *Journal of applied physics* **2002,** 92, (5), 2534-2539.
31. He, J.; Girard, S. N.; Kanatzidis, M. G.; Dravid, V. P. *Advanced Functional Materials* **2010,** 20, (5), 764-772.
32. Slack, G. A.; Galginaitis, S. *Physical Review* **1964,** 133, (1A), A253.
33. Tritt, T. M., *Thermal conductivity: theory, properties, and applications*. Springer Science & Business Media: 2004.
34. Lo, S. H.; He, J.; Biswas, K.; Kanatzidis, M. G.; Dravid, V. P. *Advanced Functional*

*Materials* **2012,** 22, (24), 5175-5184.
35. Kazan, M.; Guisbiers, G.; Pereira, S.; Correia, M.; Masri, P.; Bruyant, A.; Volz, S.; Royer, P. *Journal of Applied Physics* **2010,** 107, (8), 083503.
36. Morelli, D.; Heremans, J.; Slack, G. *Physical Review B* **2002,** 66, (19), 195304.
37. Volz, S. G.; Chen, G. *Physical Review B* **2000,** 61, (4), 2651.
38. Klemens, P. *Solid state physics* **1958,** 7, 1-98.
39. Duda, J. C.; English, T. S.; Piekos, E. S.; Soffa, W. A.; Zhigilei, L. V.; Hopkins, P. E. *Physical Review B* **2011,** 84, (19), 193301.
40. Dove, M. T., *Introduction to lattice dynamics*. Cambridge university press: 1993; Vol. 4.

# Supporting Information for
# Thermal conductivity reduction by 60° shuffle-set dislocation arrays embedded in silicon nano-films

Zhimin Jiang, Haisheng Fang, Zhenya Lv, Xiangang Luo, Lili Zheng

## 1. Nonequilibrium molecular dynamics simulation method

In MD simulations performed by LAMMPS package,[1] Stillinger-Weber (SW) potential for silicon is chosen to drive atoms.[2] It has been successfully applied to investigate thermal properties of both nanostructure and bulk silicon.[3, 4] To eliminate strains, as well as to obtain an equilibrium state, the NPT assemble is applied for 0.6 ns with a pressure $P$=0 Pa and a MD temperature[5] $T_{MD}$= 300 K given by

$$\frac{3}{2}Nk_BT_{MD} = \sum_{i=1}^{N}\frac{1}{2}m_iv_i^2, \qquad (1)$$

where $N$, $m_i$, $v_i$ and $k_B$ are the number of atoms, the atom mass, the atom velocity and the Boltzmann's constant, respectively. Although quantum corrections (QCs) have been proposed to estimate the real temperature, $T$,[6, 7] it does not result in a better agreement with quantum predictions compared to the uncorrected values above temperatures of 200K.[7, 8] In the current study, $T=T_{MD}$ is applied for approximation. In the following 0.6 ns, a stationary temperature gradient is established by the Nose-Hoover thermal baths with the temperatures of $T_{hot}$ and $T_{cold}$ (see Fig. 1 in the paper).[5] Afterwards, the added and extracted energies ($E_{hot}$ and $E_{cold}$) from the regions, as well as velocities of all of the atoms, are recorded for 3 ns. The obtained thermal conductivity is calculated via

$$\kappa = -\frac{|J_{hot}|+|J_{cold}|}{2A \times \nabla T}, \qquad (2)$$

where $A$, $J$ and $\nabla T$ are the across section area of the simulation region, the heat flow and the temperature gradient, respectively. The average heat flux $(|J_{hot}|+|J_{cold}|)/2A$ is linearly fitted from the average recorded energy per area, $(|E_{hot}|+|E_{cold}|)/2A$, as shown in Fig. S1. $\nabla T$ is linearly fitted from the calculated temperature profile (see Fig. S3), or determined by $(T_2-T_1)/(L_{21})$ (see Fig. S2), depending on the situation whether dislocations are embedded in films or not. $T_1$ and $T_2$ are the average temperatures of the third atom layers away from the hot and the cold thermal baths, respectively. $L_{21}$ is the distance between the two layers (about 9.7 nm), as shown in Fig. S2. The final $\kappa$ is averaged from four runs with different initial velocity distributions.

Before embedding dislocations in films, $\kappa$ of perfect films with different simulation dimensions in the x and the y axes are computed to test simulation-dimension effect. The results are presented in Fig. S4. Because of periodic boundaries, the effect is negligible in the x direction if x-direction size is larger than 4.0 nm. For the y directions, the situation is nearly the same. To achieve a balance between simulation accuracy and computational load, the size of simulation box along the y direction is determined to be 3.08 nm. Also, the results disclose that the perfect thin films with different simulation dimensions have similar $\kappa$ values of about 18.0 W/(mK), which is slightly higher than the results in Ref. 9 before QCs.

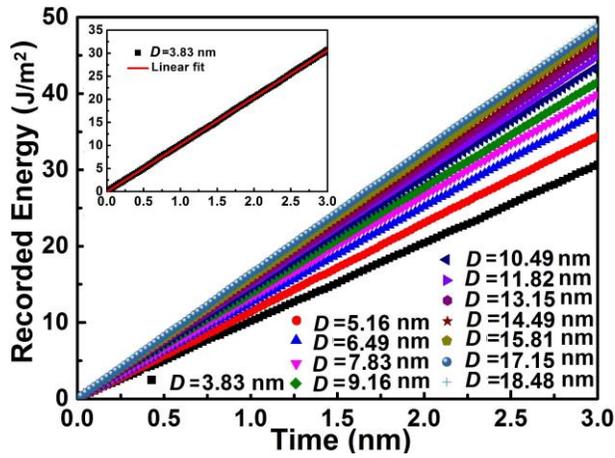

Fig. S1. The average recorded energy per area with different dislocation distances ($D$). The inset shows the linear fitting curve (red solid line) of the data with $D$=3.83 nm, of which the slope is the heat flux.

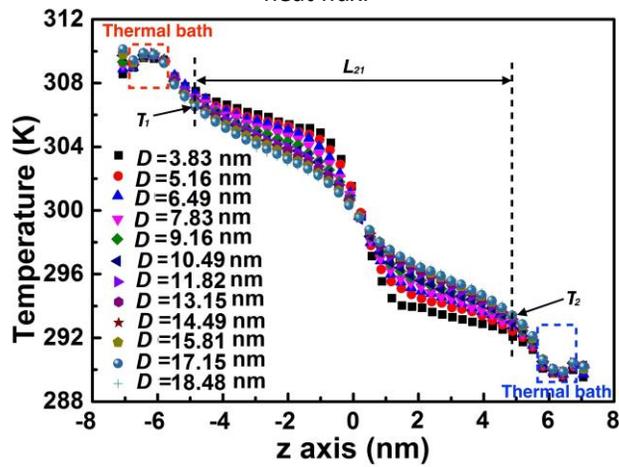

Fig. S2. Average temperatures of atom layers along z axis. Different symbols stand for different dislocation distances.

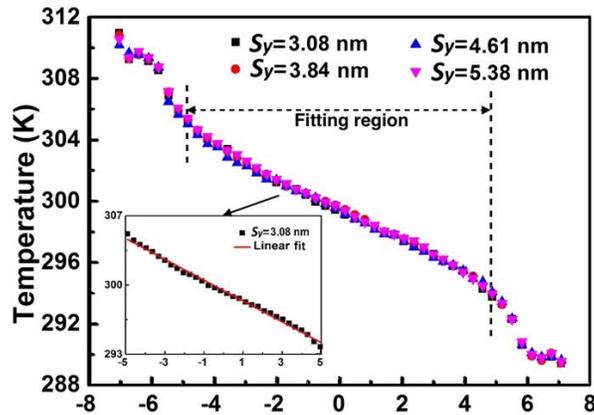

Fig. S3. Temperature distributions of perfect nano-films with different simulation sizes in the y direction ($S_y$). The red linear-fitting line in the inset is an example to illustrate the calculation of temperature gradient.

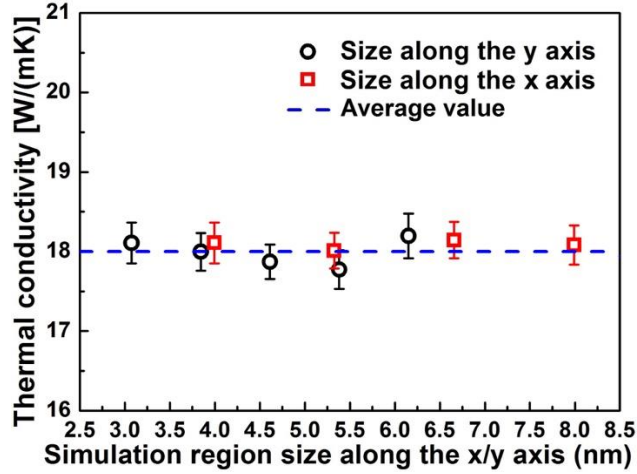

Fig. S4. The calculated κ of perfect films with different simulation dimensions in the x and the y axes. The x-axis dimension of black circles has the same value of about 4.00 nm, and the y-axis dimension of red squares is about 3.08 nm.

## 2. Debye-Callaway's and Klemens' models

A modified Debye-Callaway's model[10] is adopted to analyze the possible mechanisms for reduction of thermal conductivity in dislocation-embedded films. It is assumed that the dominant contribution to thermal conductivity is from acoustical phonons with their dispersion in the vibrational spectrum neglected. The material is further assumed isotropic. The model contains two parts[10-12] regarding $\kappa_1$ and $\kappa_2$ ($\kappa=\kappa_1+\kappa_2$) as expressed by

$$\kappa_1 = \frac{1}{3}\sum_n \frac{k_B}{2\pi^2 v_n}\left(\frac{k_B T}{\hbar}\right)^3 \int_0^{\Theta_n/T} \tau_{c,n}(x)\frac{x^4 e^x}{(e^x-1)^2}dx \tag{3a}$$

$$\kappa_2 = \frac{1}{3}\sum_n \frac{k_B}{2\pi^2 v_n}\left(\frac{k_B T}{\hbar}\right)^3 \frac{\left[\int_0^{\Theta^i/T} \tau_{c,n}(x)\tau_{N,n}(x)^{-1} x^4 e^x (e^x-1)^{-2} dx\right]^2}{\int_0^{\Theta^i/T} \tau_{c,n}(x)\left[\tau_{N,n}(x)\tau_{u,n}(x)\right]^{-1} x^4 e^x (e^x-1)^{-2} dx}. \tag{3b}$$

In Eqs.3a and 3b, the subscript n represents three acoustical phonon branches, two transverse (T) branches and one longitudinal (L) branch. $k_B$, $\hbar$ and Θ are the Boltzmann's constant, the Planck's constant, and the Debye temperature, respectively. υ is the phonon-group velocity at the low frequency limit.[11] The Debye model has a maximum phonon frequency, $\omega_D$, and a corresponding Debye temperature calculated by $k_B\Theta=\hbar\omega_D$.[6, 13] The parameter x is defined by $x=\hbar\omega/k_B T$, where ω is the angular frequency of phonons. $\tau_c$ is a combined relaxation time with $\tau_c^{-1}=\tau_N^{-1}+\tau_u^{-1}$, where $\tau_N$ and $\tau_u$ are the relaxation time for the normal process and for the processes characterized by unconserved momentum (u process), respectively. The u process includes Umklapp phonon-phonon scattering, defect scattering, boundary scattering and so on. In the current study, the scattering mechanisms result in

$$\tau_{u,n}^{-1} = \tau_{U,n}^{-1} + \tau_{SD,n}^{-1} + \tau_{ED,n}^{-1} + \tau_{DC,n}^{-1} + \tau_{B,n}^{-1}, \tag{4}$$

where $\tau_U$, $\tau_{SD}$, $\tau_{ED}$, $\tau_{DC}$ and $\tau_B$ are the relaxation times result from phonon scattering by the Umklapp process, by the screw dislocation component, by the edge dislocation component, by the dislocation cores, and by the boundaries,[10, 14] respectively. All of the relaxation times are calculated from the expressions listed in Table I.

Based on the Leibfried and Schlömann model, the relaxation time of Umklapp scattering proposed by Slack et al. follows the form of

$$\tau_U^{-1} \approx \frac{\hbar \gamma^2}{M \upsilon^2 \Theta} \omega^2 T \exp\left(\frac{-\Theta}{3T}\right), \quad (5)$$

where $M$ is the averaged atomic mass, and $\gamma$ is the Gruneisen anharmonicity parameter [15-17]. For normal process, its relaxation time is taken to be

$$\tau_N^{-1} \approx B_N \omega^a T^b, \quad (6)$$

where ($a$, $b$) equals to (1,4) for longitudinal phonons and to (2, 3) for transverse phonons, and $B_N$ is a constant independent of $T$ and $\omega$.[10, 12] But, in practice, we adopt another simpler expression, $\tau_{N,n}^{-1}=1.5\tau_{u,n}^{-1}$.[18] As for the boundary scattering, it is described by a constant,

$$\tau_B^{-1} = \frac{\upsilon}{L}, \quad (7)$$

where $L$ is the Characteristic length[10, 18] determined by fitting experiment or simulation results.

Klemens' formulas based on second-order perturbation theory for phonon scattering rates due to screw dislocations, edge dislocations and dislocation cores are written as[14, 18]

$$\tau_{SD}^{-1} = A \times N_D B_{SD}^2 \gamma^2 \omega, \quad (8)$$

$$\tau_{ED}^{-1} = A \times N_D B_{ED}^2 \gamma^2 \omega \left\{ \frac{1}{2} + \frac{1}{24}\left(\frac{1-2v}{1-v}\right)^2 \left[1+\sqrt{2}\left(\frac{\upsilon_L}{\upsilon_T}\right)^2\right]^2 \right\}, \quad (9)$$

and

$$\tau_{DC}^{-1} = N_D \times \frac{3V_0^{4/3} \omega^3}{4\upsilon^2} \times \left(\frac{\Delta M}{M}\right)^2, \quad (10)$$

where $N_D$, $V_0$, and $v$ are dislocation density, atomic volume and Poisson's ratio.[11, 12, 14, 18] $B_{SD}$ and $B_{ED}$ are the screw and the edge components of Burgers' vector obeying $B^2 = B_{SD}^2 + B_{ED}^2$. The equation 8 and 9 were first proposed by Klemens with A=0.06 for dislocations perpendicular to temperature gradient.[14] In 1958, Klemens pointed out A should be multiplied by a factor of 16,[19] which was applied in investigations of thermal resistance in alloys.[20] According to his latest book and other researches, we adopted the value of 1.1 for our studies.[18, 21]

Before applying the models to nanostructures, the models should be fitted to the published experimental and MD simulation results of bulk pure silicon. The fitted curve and the related parameters are shown in Fig. S5 and Table II. In the table, $\upsilon_L$ and $\upsilon_T$ approximately equals to group velocities of each branch at the gamma point.[22] $\Theta_L$ and $\Theta_T$ are deduced from $\omega_{D,L}$=11.8 THz and $\omega_{D,T}$=4.5 THz, respectively, with regards to phonon dispersion curve[22]. The only adjustable parameters are $L$ and $\gamma$, among which the latter is chosen to be close to Ref. 23. For bulk silicon, $L$ is about 14 mm, fitted by thermal conductivity at temperatures close to 0 K. For thin silicon films, by fitting to the current MD simulation results of perfect films at 300 K (see Fig. S4), $L$ is determined to be about 43 nm, which is slightly smaller than mean free path of silicon nanowires (about 60 nm).[4] Because of extreme small size, $\kappa$ of nano-films or nanowires are mainly limited by the boundary scattering, resulting in $L$ approximately equal to mean free path of phonons. Then, dislocation scattering terms are added in the Callaway's model, and their parameters adopted to predict $\kappa$ of dislocation-embedded films are listed in Table III, where the only adjustable parameter is $\Delta M/M$.

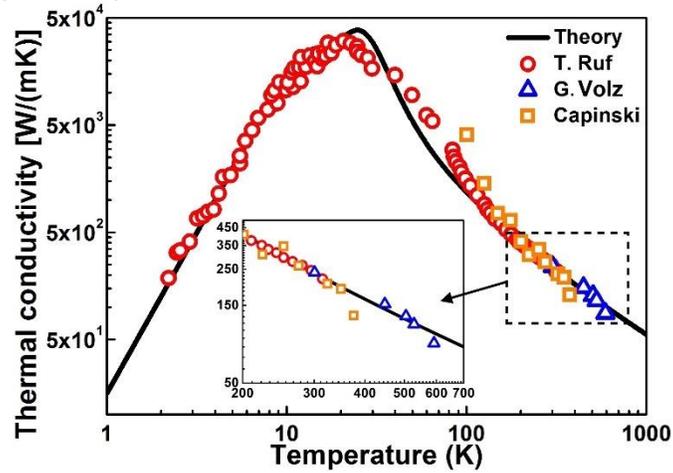

Fig. S5 The fitted curve compared to the experimental and MD simulation results.[24-26]

## 2. Calculation method for local phonon density of states

For investigation of phonon scattering, local phonon density of states (LPDOS), as a density distribution function of phonon eigenstates of each atom, is a critical parameter. Its consistence degree between the atoms represents the smooth degree of the phonon transport. The LPDOS is proportional to the Fourier transform of the velocity autocorrelation function, but, in practice, is usually calculated by the standard estimation procedures for power spectral density, i.e., the Welch method.[27] The atom velocities around the dislocation are tracked for 655.36 ps with the NVE assemble at $T_{MD}$=300 K. The velocity series of each atom are divided into 15 segments with 50% overlapping with each neighboring segment. After multiplied by the Hamming window, each segment is then transformed into the frequency domain by the Fourier transform. Finally, the normalized LPDOS around the dislocation are obtained by normalizing the squared averaged magnitudes of the Fourier-transformed segment of each atom.

**Reference**
1.   Plimpton, S. *Journal of computational physics* **1995,** 117, (1), 1-19.
2.   Stillinger, F. H.; Weber, T. A. *Physical review B* **1985,** 31, (8), 5262.
3.   Volz, S. G.; Chen, G. *Applied Physics Letters* **1999,** 75, (14), 2056-2058.
4.   Yang, N.; Zhang, G.; Li, B. *Nano Today* **2010,** 5, (2), 85-90.
5.   Evans, D. J.; Holian, B. L. *The Journal of chemical physics* **1985,** 83, (8), 4069-4074.
6.   Dove, M. T., *Introduction to lattice dynamics*. Cambridge university press: 1993; Vol. 4.

7. Kaviany, M., *Heat transfer physics*. Cambridge University Press: 2014.
8. Turney, J.; McGaughey, A.; Amon, C. *Physical Review B* **2009,** 79, (22), 224305.
9. Gomes, C. J.; Madrid, M.; Goicochea, J. V.; Amon, C. H. *Journal of heat transfer* **2006,** 128, (11), 1114-1121.
10. Callaway, J. *Physical Review* **1959,** 113, (4), 1046.
11. Zou, J.; Kotchetkov, D.; Balandin, A.; Florescu, D.; Pollak, F. H. *Journal of applied physics* **2002,** 92, (5), 2534-2539.
12. He, J.; Girard, S. N.; Kanatzidis, M. G.; Dravid, V. P. *Advanced Functional Materials* **2010,** 20, (5), 764-772.
13. Toberer, E. S.; Zevalkink, A.; Snyder, G. J. *Journal of Materials Chemistry* **2011,** 21, (40), 15843-15852.
14. Klemens, P. *Proceedings of the Physical Society. Section A* **1955,** 68, (12), 1113.
15. Slack, G. A.; Galginaitis, S. *Physical Review* **1964,** 133, (1A), A253.
16. Tritt, T. M., *Thermal conductivity: theory, properties, and applications*. Springer Science & Business Media: 2004.
17. Lo, S. H.; He, J.; Biswas, K.; Kanatzidis, M. G.; Dravid, V. P. *Advanced Functional Materials* **2012,** 22, (24), 5175-5184.
18. Kim, S. I.; Lee, K. H.; Mun, H. A.; Kim, H. S.; Hwang, S. W.; Roh, J. W.; Yang, D. J.; Shin, W. H.; Li, X. S.; Lee, Y. H. *Science* **2015,** 348, (6230), 109-114.
19. Klemens, P. *Solid state physics* **1958,** 7, 1-98.
20. Lomer, J. N.; Rosenberg, H. *Philosophical Magazine* **1959,** 4, (40), 467-483.
21. Klemens, P. *Academic (New York, 1969) p* **1969,** 1.
22. Kazan, M.; Guisbiers, G.; Pereira, S.; Correia, M.; Masri, P.; Bruyant, A.; Volz, S.; Royer, P. *Journal of Applied Physics* **2010,** 107, (8), 083503.
23. Morelli, D.; Heremans, J.; Slack, G. *Physical Review B* **2002,** 66, (19), 195304.
24. Ruf, T.; Henn, R.; Asen-Palmer, M.; Gmelin, E.; Cardona, M.; Pohl, H.-J.; Devyatych, G.; Sennikov, P. *Solid State Communications* **2000,** 115, (5), 243-247.
25. Capinski, W.; Maris, H.; Bauser, E.; Silier, I.; Asen-Palmer, M.; Ruf, T.; Cardona, M.; Gmelin, E. *Applied physics letters* **1997,** 71, (15).
26. Volz, S. G.; Chen, G. *Physical Review B* **2000,** 61, (4), 2651.
27. Duda, J. C.; English, T. S.; Piekos, E. S.; Soffa, W. A.; Zhigilei, L. V.; Hopkins, P. E. *Physical Review B* **2011,** 84, (19), 193301.